\documentclass[twocolumn,showpacs,preprintnumbers,amsmath,amssymb,showkeys]{revtex4}
\usepackage{graphicx}% Include figure files
\usepackage{dcolumn}% Align table columns on decimal point
\usepackage{bm}% bold math

\begin{document}
\preprint{APS/123-QED}

\title{Cluster structure of a low-energy resonance in tetraneutron}
\author{Yuliya Lashko}
\email{lashko@univ.kiev.ua}
\author{Gennady Filippov}
\email{gfilippov@bitp.kiev.ua}\affiliation{Bogolyubov Institute
for Theoretical Physics\\ 14-b Metrolohichna Str., Kiev-143,
Ukraine}
\date{\today}

\begin{abstract}
We theoretically investigate the possibility for a tetraneutron to
exist as a low-energy resonance state. We explore a microscopic
model based on the assumption that the tetraneutron can be treated
as a compound system where $^3$n+n and $^2$n+$^2$n coupled cluster
configurations coexist. The influence of the Pauli principle on
the kinetic energy of the relative motion of the neutron clusters
is shown to result in their attraction. The strength of such
attraction is high enough to ensure the existence of a low-energy
resonance in the tetraneutron, provided that the oscillator length
is large enough.
\end{abstract}

%Use showkeys class option if keyword
% 21.60.-n Nuclear structure models and methods
%21.60.Gx Cluster models
%21.45.+v Few-body systems
%25.60.-t Reactions induced by unstable nuclei

\maketitle

The first claim of the experimental observation of the nuclear
stable tetraneutron has been made in \cite{Ageev}. Since then all
other experimental attempts to find either a bound or a resonance
state in the system of four neutrons have not met with success.
However, in a recently reported experiment with the breakup of
$^{14}$Be \cite{Marques2002} six events consistent with the
formation of a bound tetraneutron were revealed. Unfortunately,
several other experiments \cite{Bouchat,Rich,Aleksandrov}
undertaken to verify these results failed to prove or refute
completely the existence of the tetraneutron due to a poor
statistics. An overall conclusion of a number of theoretical
papers on this subject \cite{Pieper,Timofeyuk,Lazauskas,Nesterov}
is that it does not seem possible to change modern nuclear
Hamiltonians to bind a tetraneutron without destroying many other
successful predictions of those Hamiltonians. For instance,
calculations within the hyperspherical functions method (HSFM)
\cite{Timofeyuk} suggest that a very strong phenomenological
four-nucleon force is needed in order to bind the tetraneutron.
And yet neither theoretical nor experimental  results exclude the
possible existence of tetraneutron as a low-energy resonance (see
\cite{Pieper,Nesterov,Marques2005}).

There are only few cases of theoretical treatment of the resonant
tetraneutron. In Ref. \cite{Nesterov} the continuum states of the
$^4$n system were studied in the framework of the approach which
combines concepts of the HSFM and the resonating-group method
(RGM). Along with the lowest order hyperharmonic the authors of
Ref. \cite{Nesterov} invoked the hyperharmonics with hypermomenta
$K=K_{\rm{min}}+2$ and $K=K_{\rm{min}}+4$, which reproduce
$^{2}$n+$^{2}$n clustering of the tetraneutron. The analysis of
the energy behaviour of the eigenphases led authors to the
conclusion that $^4$n has a resonance state at an energy of about
1-3 MeV. But clear indication of such a resonance was obtained
only for the Volkov effective $NN$ potential, which is known to be
inappropriate for studying multineutron systems as it binds
dineutron.

The most systematic study of four-neutron resonances was
performed in \cite{Lazauskas}, where configuration space
Faddeev-Yakubovsky equations have been solved using realistic $NN$
interaction to follow the resonance pole trajectories in the
complex energy plane. It was concluded in \cite{Lazauskas} that
tetraneutron –-- bound or resonant –-- can be created only in
strong external fields and would disintegrate right after such a
field is removed. The same authors remarked, however, that an
accurate determination of the physical resonance position was not
possible with the methods used in \cite{Lazauskas}.

Hence, the existence of resonant tetraneutrons is still an open
question. In the present paper, we study the possibility for the
tetraneutron to be described as a compound system where $^3$n+n
and $^2$n+$^2$n coupled cluster configurations coexist.

 The Pauli exclusion
principle is known to significantly influence the interaction of
composite particles. An exact treatment of the antisymmetrization
effects related to the kinetic energy exclusively was shown to
result in either an effective repulsion or attraction of the
clusters \cite{PRC}. Such an effective interaction substantially
affects the dynamics of the cluster-cluster interaction and can,
on occasion, produce resonance behaviour of the scattering phase
or even a bound state in compound nuclear system. The role of the
Pauli principle in the formation of both the discrete spectrum and
multi-channel states of the two-cluster nuclear systems can be
studied in the Algebraic Version of the resonating-group method
(AVRGM) \cite{PEPAN}. An RGM wave function is built in the form of
an antisymmetrized product of cluster internal wave functions
\footnote{They are assumed to be fixed and described by the
translation-invariant shell model wave functions with the same
oscillator length $r_0$ for both clusters.} and a wave function of
their relative motion. The latter depends only on the Jacobi
vector ${\bf R}$ of the considered binary system and, according to
the AVRGM, is expanded in a complete discrete basis of
harmonic-oscillator (HO) states allowed by the Pauli principle
\begin{eqnarray*}
\Psi_{\kappa\,(E)}({\bf R})=\sum_n
\sqrt{\Lambda_n}C_n^{\kappa\,(E)}\psi_n({\bf R}).
\end{eqnarray*}
The Pauli-allowed basis functions $\psi_n$ are defined in the
Fock–-Bargmann representation \cite{Barg}, where they take
exceptionally simple form. Expansion coefficients both of the
discrete eigenstates with energy $E_\kappa=-\kappa^2/2<0$ and of
the continuum eigenstates $\left\{C_{n}(E)\right\}$ with energy
$E>0$ are found by solving a set of linear equations
\begin{eqnarray}
\label{eq} \sum_{n'}\langle n|\hat{H}|n'\rangle C_{n'}-E
\Lambda_nC_{n}=0.
\end{eqnarray}
The asymptotic behavior of the continuum eigenstates is expressed
in terms of Hankel functions of the first and second kind, and
the scattering $S$-matrix elements.

The set of quantum numbers $n$ includes the number of oscillator
quanta $\nu$, the indices $(\lambda,\mu)$ of their SU(3)
symmetry, the additional quantum number $\alpha_{(\lambda,\mu)}$
if there are several differing $(\lambda,\mu)$ multiplets, the
orbital momentum $L$ and its projection $M$, and one more
additional quantum number $\alpha_L$ if the multiplet
$(\lambda,\mu)$ has several states with the same values of $L$.
In this paper we restrict the discussion to the case of $L=M=0$
and, hence, we will use only first three quantum numbers.

The influence of the Pauli exclusion principle on the collision
of clusters through the kinetic energy is determined by the
eigenvalues $\Lambda_\nu$ of the antisymmetrizer:
\begin{eqnarray*}
\hat{A}\psi_{\nu}=\Lambda_\nu\psi_\nu,~~\Lambda_\nu\geq0,~~\Lambda_\nu\rightarrow1,
\nu\rightarrow\infty.
\end{eqnarray*}
The eigenvalues are proportional to the probability of the system
being in the state determined by the eigenfunction $\psi_\nu$.
$\Lambda_\nu>0$ correspond to the Pauli-allowed states, while
zero eigenvalues belong to the Pauli-forbidden states.
$\Lambda_\nu<1$ generate repulsion of clusters, whereas an
attraction appears in the states with the eigenvalues exceeding
unity.

Due to the exchange of nucleons belonging to different clusters,
Eq. (\ref{eq}) resembles the Schr\"{o}dinger equation for relative
motion of two particles with highly nonlocal interaction between
them. This complicates an analysis of the cluster-cluster
interaction generated by the nucleon-nucleon forces. However, if
the oscillator length $r_0$ is much less than the range $b_0$ of
the $NN$ potential, then the potential energy matrix in the HO
representation is equivalent to the diagonal matrix which is a
discrete analog of cluster-cluster potential in the coordinate
space
\begin{eqnarray}
\label{pot} \sum_{\nu'_{\min}}^\infty \langle\nu,l|U|\nu',l\rangle
C_{\nu',l}=U_{\rm{eff}}(r_\nu) C_{\nu,l}.
\end{eqnarray}
For a central nucleon-nucleon potential having a Gaussian form we
will have
\begin{equation}
\label{pot_as} U_{\rm{eff}}(r_\nu)\sim
U_0\exp\left\{-\frac{r_\nu^2}{b_0^2}\right\},
r_\nu=\sqrt{\frac{A_1+A_2}{A_1A_2}}r_0\sqrt{2\nu+2l+3},
\end{equation}
where $U_0$ is the strength of the Gaussian potential, $r_\nu$
defines the distance between clusters (composed of $A_1$ and $A_2$
nucleons, respectively) and $l$ is the angular momentum of the
cluster relative motion.

\begin{figure}[tbh]
\includegraphics[width=8.6cm]{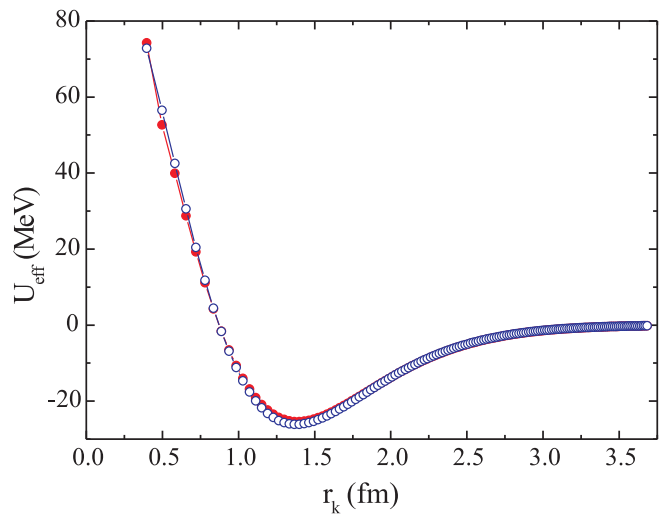}
\caption{\label{fig:1} (Color online) Effective $^2n$-$^2n$
potential generated by the Serber interaction. Solid circles:
$\sum_{k'=1}^{\infty}\langle k|U|k'\rangle$. Empty circles:
equivalent local potential $U_{\rm{eff}}(r_k)$ (see text for
details).}
\end{figure}

\begin{figure}[tbh]
\includegraphics[width=8.6cm]{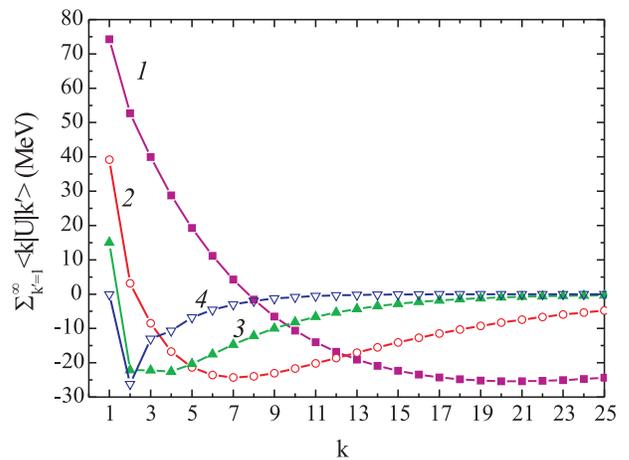}
\caption{\label{fig:2} (Color online) Effective $^2n$-$^2n$
potential generated by the Serber interaction versus the number of
quanta $k$ at different values of oscillator length $r_0$. Curves:
\textit{1} -- $r_0=0.15$ fm; \textit{2} -- $r_0=0.25$ fm;
\textit{3} -- $r_0=0.35$ fm; \textit{4} -- $r_0=0.5$ fm.}
\end{figure}

The validity of relation (\ref{pot}) is demonstrated in
Fig.~\ref{fig:1} by the example of effective $^2n$-$^2n$ potential
generated by the Serber $NN$ interaction \cite{Serber} ($r_0=0.15$
fm, $b_0$=1.48 fm, $\nu=2k$).
So, we deduce from (\ref{pot_as}) that the effective
cluster-cluster potential decreases exponentially as oscillator
length $r_0$ increases (see Fig.~\ref{fig:2}).
It would appear natural that $r_0$ should be large enough to give
a reasonable fit to the neutron density distribution inside
dineutron clusters, i.e. $r_0>b_0$. But then we immediately arrive
at the conclusion that \textit{effective intercluster interaction
derived from the $NN$ potential is of minor importance and can not
produce a resonance in $^4n$ system}. We have used the
Gaussian-type potential for simplicity, but all the above is
certain to remain valid for any short-range potential.

\begin{figure}[tbh]
\includegraphics[width=8.6cm]{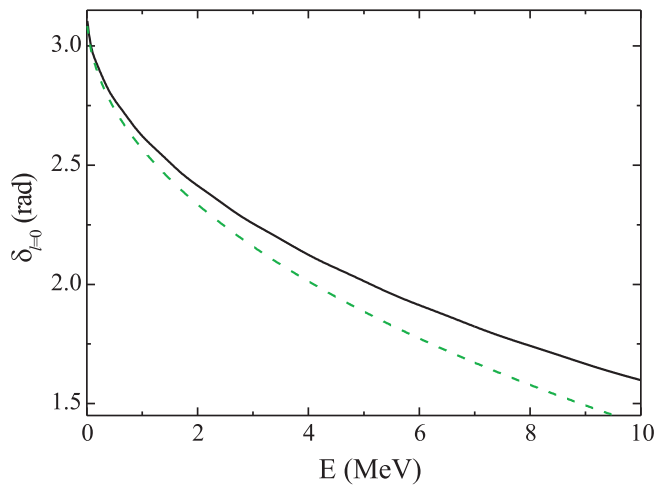}
\caption{\label{fig:3} (Color online) Phase shift of the $^2n+^2n$
scattering, $r_0=b_0=1.48$ fm. Dashed curve: the phase obtained by
incorporating the kinetic energy exchange potential exclusively.
Solid curve: phase shift calculated with the inclusion of the
cluster-cluster potential generated by Ripka forces \cite{Ripka}.}
\end{figure}

As regards the contribution from the kinetic energy operator (with
its exchange part), in this case it is reduced to the elimination
of a single Pauli forbidden state by virtue of the fact that for
the $^2n+^2n$ configuration the eigenvalues of all the allowed
states equal to unity. Thus, the eigenfunctions of the kinetic
energy operator in the region of relatively small energies are
similar to the eigenfunctions of a particle in the field of a hard
core, and, in the region of large energies, they are identical to
the wave functions of free motion of a particle. Hence, there is
no grounds to believe that the tetraneutron can exist as a
dineutron-dineutron cluster system. Behaviour of the phase shift
of the $^2n+^2n$ scattering, shown in Fig.~\ref{fig:3}, supports
this inference.
Figure~\ref{fig:3} also corroborates our conclusion that the
$^2$n-$^2$n interaction generated by the $NN$ forces has a minor
effect on physical observables \footnote{We used the Ripka $NN$
force to accentuate that even purely attractive $NN$ potential is
not able to contribute significantly to the scattering phase.}.

For each compound nuclear system, one can identify several
cluster configurations corresponding to a certain set of clusters
released as a result of a nuclear reaction. The probability for a
cluster configuration to be realized in the Pauli-allowed basis
function of a binary cluster system with the minimum number of
quanta was shown to be proportional to the eigenvalue of isolated
configuration \cite{YaF06}. Maximum eigenvalues belong to the
$^3$n+n configuration of the $^4$n system. They exceed unity and
have a maximum at the minimum number of quanta that can be
interpreted as an attraction of neutron clusters at small
distance between them. Taking into account various cluster
configurations of the compound nucleus causes the maximum
eigenvalues to increase. Moreover, new branches of excitation
appear with particularly large (larger than unity) eigenvalues of
the allowed states that is an evidence of essential attraction of
the clusters due to exchange effects. Therefore, we have good
reason to believe that consideration of $^2n+^2n$ and $^3n+n$
coupled cluster configurations of the $^4n$ system will allow us
to solve the problem of existence of a bound state or a resonance
in the tetraneutron.

\begin{figure}[tbh]
\includegraphics[width=8.6cm]{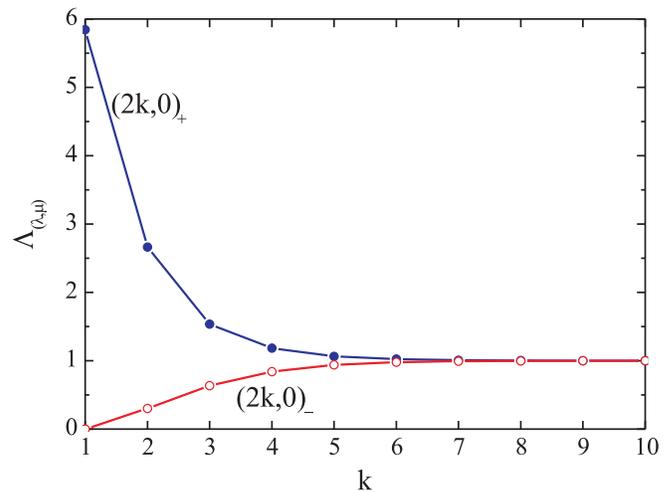}
\caption{\label{fig:4} (Color online) Eigenvalues
$\Lambda_{(\lambda,\mu)}$ of the allowed states for the $^4n$
system in the coupled-channel approach.}
\end{figure}

\begin{figure}[tbh]
\includegraphics[width=8.6cm]{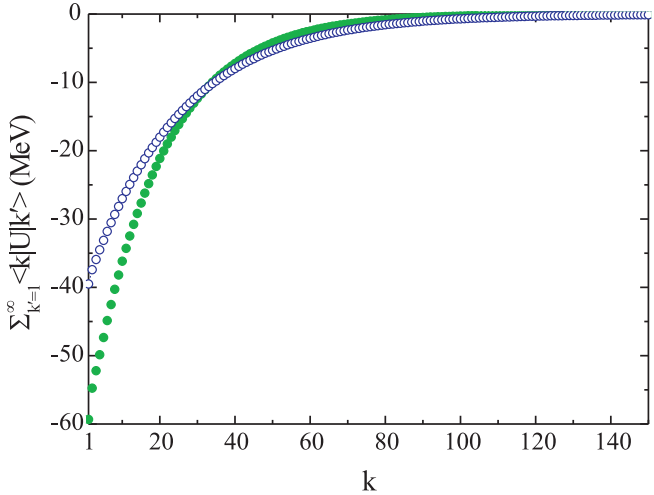}
\caption{\label{fig:5} (Color online) Effective intercluster
potential induced by the Ripka interaction. Solid and empty
circles: $^3$n+n and $^2$n+$^2$n cluster configurations,
correspondingly.}
\end{figure}

The $^2n+^2n$ and the $^3n+n$ configuration coupled provide two
branches of the two-fold degenerate SU(3) representation
$(2k,0)$. The behaviour of the eigenvalues belonging to these
branches is shown in Fig.~\ref{fig:4}. An effective attraction
reveals itself in $(2k,0)_+$-branch while in $(2k,0)_-$-branch an
effective repulsion takes place.
The remarkable feature of these branches is that at large
distance between clusters $(k\rightarrow\infty)$ they contain the
wave function of both cluster configurations on equal terms:
\begin{eqnarray*}
\Psi_{(2k,0)_\pm}\rightarrow{1\over\sqrt{2}}
\left\{\psi_{(2k,0)}(^3n+n)\pm\bar{\psi}_{(2k,0)}(^2n+^2n)\right\}.
\end{eqnarray*}
Returning to the analysis of intercluster interaction generated
by $NN$ forces, we refer to Fig.~\ref{fig:5}.
Figure ~\ref{fig:5} depicts the effective cluster-cluster
potentials generated by the Ripka forces and calculated for
isolated cluster configurations of the $^4$n system. It is seen
from Fig.~\ref{fig:5} that the strength of $^3$n-n interaction
exceeds that of $^2$n-$^2$n interaction by the difference
$\Delta_u$ in intrinsic potential energy of the $^3$n cluster and
two dineutrons. For the Gaussian-type potential
$\Delta_u=-0.5V_{13}(1+2r_0^2/b_0^2)^{-3/2},$ where $V_{13}$ is
the strength of interaction of a neutron pair in singlet spin
state.

\begin{figure}[tbh]
\includegraphics[width=8.6cm]{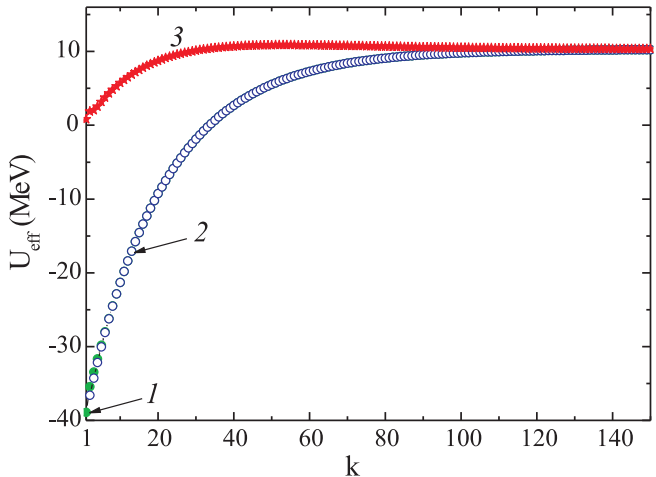}
\caption{\label{fig:6} (Color online) Effective cluster-cluster
potential $U_k^{\rm{eff}}$ generated by Ripka interaction for the
$^4n$ system in the coupled-channel approach. Curves: \textit{1}
-- $U_k^{++}=\sum_{k'=1}^{\infty}\langle
(2k,0)_+|U|(2k',0)_+\rangle$; \textit{2} --
$U_k^{--}=\sum_{k'=1}^{\infty}\langle
(2k,0)_-|U|(2k',0)_-\rangle$; \textit{3} --
$U_k^{+-}=\sum_{k'=1}^{\infty}\langle
(2k,0)_+|U|(2k',0)_-\rangle$.}
\end{figure}

\begin{figure}[tbh]
\includegraphics[width=8.6cm]{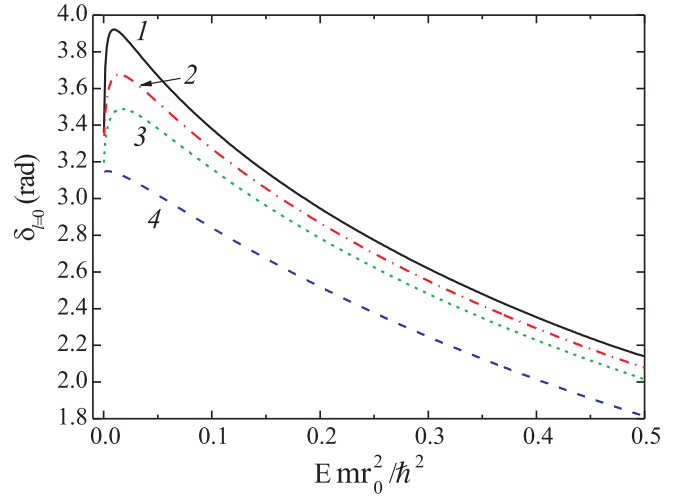}
\caption{\label{fig:7} (Color online) Phase shift of the elastic
$^2n+^2n$ scattering in the coupled-channel approach, provided
that $\Delta_u=0$ (\textit{1}) and $\Delta_u\neq0$ (produced by
Ripka nucleon-nucleon forces of the range $b=1.48$ fm) at
$r_0=10b$ (\textit{2}); $r_0=5b$ (\textit{3}) and $r_0=b$
(\textit{4}).}
\end{figure}

The $\Delta_u$ causes the SU(3)-branches $(2k,0)_+$ and
$(2k,0)_-$ to remain coupled even at large $k$ (see
Fig.~\ref{fig:6}).
The difference in intrinsic potential energies $\Delta_u$ of
$^3n$  and two dineutrons, along with the difference in their
kinetic energies $\Delta_t$, gives rise to the threshold of the
tetraneutron decay into $^3n$ and a neutron. Although the
strength of such energy barrier $\Delta=\Delta_u+\Delta_t$ is
mainly determined by the discrepancy between intrinsic kinetic
energies of the clusters \footnote{$\Delta_t=0.5\hbar^2/mr_0^2$,
$m$ denotes nucleon mass.}, the latter does not couple basis
functions belonging to different SU(3) branches. At the same
time, nonzero $\Delta_u$ implies additional repulsion, which
prevents the formation of resonance in the tetraneutron.
$\Delta_u$ tends to diminish with increasing the oscillator
length $r_0$. Hence, \textit{effective attraction generated by the
kinetic exchange terms is able to create a low-energy resonance
in $^4$n only at large values of $r_0$}. Figure~\ref{fig:7}, which
represents the phase shift of the elastic $^2n+^2n$ scattering
versus dimensionless energy, sustains this assumption.
A resonance behaviour of the phase shift is observed only at the
values of $r_0$ of the order of 10 fm. This resonance is very
close to zero energy above the $^4$n$\rightarrow^2$n+$^2n$
threshold, and its width is quite large in comparison with its
energy. Hence, our results are consistent with the experimental
data \cite{Marques2002}. Indeed, recently the authors of the
latter paper discussed the description of these data by means of
an unbound-tetraneutron resonance in \cite{Marques2005}. They came
to recognize that a broad low-energy resonance in the tetraneutron
can be compatible with the events observed in \cite{Marques2002}.
Moreover, theoretical \textit{ab initio} calculations of Pieper
\cite{Pieper} also leaves room for the occurrence of a broad $^4$n
resonance with an energy of around 2 MeV or less.

In conclusion, we revealed that the tetraneutron has a good
chance to exist as a compound system where $^3$n+n and $^2$n+$^2$n
coupled cluster configurations coexist. The influence of the Pauli
principle on the kinetic energy of the relative motion of the
neutron clusters was shown to result in their attraction. The
strength of such attraction is high enough to ensure the
existence of a low-energy resonance in the tetraneutron, provided
that the oscillator length is large enough. It was also
demonstrated that increasing the oscillator length results in a
depression of the cluster-cluster potential. For this reason the
results are not sensitive to the choice of the phenomenological
two-body nucleon-nucleon potential. No three- and four-nucleon
forces are employed in the calculations.

We believe that our theoretical predictions can provide fresh
insight into the problem of existence of resonance states in pure
neutron systems and  may help to illuminate the nature of such
states and mechanism of their formation.


\begin{thebibliography} {99}

\bibitem{Ageev} V.~A.~Ageev \emph{et al.}, Ukr. Phys. J., {\bf 31}, 1771 (1986).
\bibitem{Marques2002} F.~M.~Marqu\'{e}s \emph{et al.}, Phys. Rev. C {\bf
65}, 044006 (2002).
\bibitem{Bouchat} V.~Bouchat \emph{et al.}, Proc. to Int. Symp.
Exotic Nuclei, Peterhof, July 5-12, 2004 (World Scientific,
2005), p. 29.
\bibitem{Rich} E.~Rich \emph{et al.}, Proc. to Int. Symp.
Exotic Nuclei, Peterhof, July 5-12, 2004 (World Scientific,
2005), p. 36.
\bibitem{Aleksandrov} D.~V.~Aleksandrov \emph{et al.}, JETP Lett.,
{\bf 81}, 43 (2005).
\bibitem{Pieper} S.~C.~Pieper, Phys. Rev. Lett. {\bf 90}, 252501 (2003).
\bibitem{Timofeyuk} N.~K.~Timofeyuk, J. Phys. G {\bf 29}, L9 (2003).
\bibitem{Lazauskas} R.~Lazauskas,J.~Carbonell, Phys. Rev. C {\bf 72}, 034003
(2005).
\bibitem{Nesterov} I.~F.~Gutich, A.~V.~Nesterov and
I.~P.~Okhrimenko, Yad. Fiz. {\bf 50}, 19 (1989).
\bibitem{Marques2005} F.~M.~Marqu\'{e}s, Eur. Phys. J. A {\bf 25}, s01,
311 (2005).
\bibitem{PRC}
Gennady Filippov and Yuliya Lashko, Phys. Rev. C \textbf{70},
064001 (2004).
\bibitem{PEPAN}
G.~F. Filippov and Yu.~A. Lashko, Phys. Part. Nucl., \textbf{36},
714 (2005).
\bibitem{Barg}
V.~Bargmann, Ann. Math. {\bf 48}, 568 (1947).
\bibitem{Serber}
D.~R.~Thompson, M.~Lemere and Y.~C.~Tang, Nucl. Phys. A
\textbf{286}, 53 (1977).
\bibitem{Ripka}
G.~Ripka, Thes. Doct. Sci. Fac. Sci. Orsay, Rapp. CEA ¹ 3404
(Univ. Paris, 1968).
\bibitem{YaF06}
Yu.~A. Lashko and G.~F. Filippov, Phys. Atom. Nucl., in press.

\end{thebibliography}
\end{document}